\documentclass[traditabstract]{aa}

\usepackage{natbib}
\usepackage{graphicx}
\usepackage{amssymb}
\usepackage{amsmath}
\usepackage{subfig}
\usepackage{caption}
\usepackage[varg]{txfonts}

\begin{document}

\newcommand{\3}{\ss}
\newcommand{\n}{\noindent}
\newcommand{\eps}{\varepsilon}
\newcommand{\be}{\begin{equation}}
\newcommand{\ee}{\end{equation}}
\newcommand{\bl}[1]{\mbox{\boldmath$ #1 $}}
\def\ba{\begin{eqnarray}}
\def\ea{\end{eqnarray}}
\def\de{\partial}
\def\msun{\mathrm{M}_\odot}
\def\msol{\mathrm{M}_\odot}
\def\te{T_{\rm eff}}
\def\logg{\log g}
\def\lmix{l_{\rm mix}}
\def\lint{l_{\rm mix}^{\rm int}}
\def\latm{l_{\rm mix}^{\rm atm}}
\def\Hp{H_{\rm P}}
\def\mv{M_{\rm V}}
\def\mi{M_{\rm I}}
\def\mj{M_{\rm J}}
\def\mk{M_{\rm K}}
\def\lsol{L_\odot}
\def\lbol{L_{\rm bol}}
\def\div{\nabla\cdot}
\def\grad{\nabla}
\def\rot{\nabla\times}
\def\ltsima{$\; \buildrel < \over \sim \;$}
\def\simlt{\lower.5ex\hbox{\ltsima}}
\def\gtsima{$\; \buildrel > \over \sim \;$}
\def\simgt{\lower.5ex\hbox{\gtsima}}

\newcommand{\cp}{\citep}
\newcommand{\ct}{\citet}
\newcommand{\cta}{\citetalias}
\newcommand{\MUSIC}{{\sf MUSIC}}
\newcommand{\ATHENA}{{\sf ATHENA}}
\newcommand{\PENCIL}{{\sf PENCIL}}

\title{Benchmarking the Multi-dimensional Stellar Implicit Code MUSIC}
\titlerunning{Benchmarking MUSIC}

  \author{T. Goffrey\inst{\ref{inst1}}
    \and J. Pratt\inst{\ref{inst1}}
    \and M. Viallet\inst{\ref{inst3}}
    \and I. Baraffe\inst{\ref{inst1},\ref{inst2}}
    \and M. V. Popov\inst{\ref{inst2}}
    \and R. Walder\inst{\ref{inst2}}
    \and D. Folini\inst{\ref{inst2}}
    \and C. Geroux\inst{\ref{inst1}}
    \and T. Constantino\inst{\ref{inst1}}}

   \institute{University of Exeter, Physics and Astronomy, EX4 4QL Exeter, UK, \email{t.goffrey@exeter.ac.uk}\label{inst1}
   \and Universit\'e de Lyon, Ens de Lyon, Univ Lyon1, CNRS, Centre de Recherche Astrophysique de Lyon UMR5574, F-69007, Lyon, France \label{inst2}
   \and Max$-$Planck$-$Institut f\"ur Astrophysik, Karl Schwarzschild Strasse 1, 85741 Garching, Germany\label{inst3}}

\authorrunning{T. Goffrey et al.}

\date{Received; accepted}

\abstract{We present the results of a numerical benchmark study for the MUlti-dimensional Stellar Implicit Code (MUSIC) based on widely applicable two- and three-dimensional compressible hydrodynamics problems relevant to stellar interiors. \MUSIC\ is an implicit large eddy simulation code that uses implicit time integration, implemented as a Jacobian-free Newton Krylov method. A physics based preconditioning technique which can be adjusted to target varying physics is used to improve the performance of the solver. The problems used for this benchmark study include the Rayleigh-Taylor and Kelvin-Helmholtz instabilities, and the decay of the Taylor-Green vortex. Additionally we show a test of hydrostatic equilibrium, in a stellar environment which is dominated by radiative effects. In this setting the flexibility of the preconditioning technique is demonstrated. This work aims to bridge the gap between the hydrodynamic test problems typically used during development of numerical methods and the complex flows of stellar interiors. A series of multi-dimensional tests are performed and analysed. Each of these test cases is analysed with a simple, scalar diagnostic, with the aim of enabling direct code comparisons. As the tests performed do not have analytic solutions we verify \MUSIC\ by comparing to established codes including \ATHENA\ and the \PENCIL\ code. \MUSIC\ is able to both reproduce behaviour from established and widely-used codes as well as results expected from theoretical predictions. This benchmarking study concludes a series of papers describing the development of the \MUSIC\ code and provides confidence in the future applications.}


\maketitle

\section{Introduction}
Despite the inherent three-dimensional nature of stellar interiors the timescales involved in stellar evolution necessitate the use of one-dimensional models. Stellar flows are multi-dimensional and non-linear in character. Therefore the one-dimensional approach requires parametrisation of three-dimensional effects. Examples of three-dimensional phenomena parametrised into one-dimensional effects are convection, through mixing length theory \citep{vitense1953,bohm1958,brandenburg2015stellar}, accretion \citep{siess1996physics,siess1997physics} and shear driven mixing \citep{zahn1992circulation,maeder1996diffusive}. With advances in current computing capability the use of multi-dimensional calculations to calibrate and improve such parametrisations is becoming increasingly feasible. Attempts to improve models of stellar convection have received considerable interest, via the so-called 321D link, \ct{arnett2015beyond}. Recent multi-dimensional tests of one-dimensional accretion models were carried out by \ct{geroux16}.

The hydrodynamical processes that influence stellar evolution are non-linear in nature, and not well represented by the idealised test problems available. Many standard test problems for compressible hydrodynamics are supersonic and dominated by shocks, and therefore not representative of the subsonic flows prevalent within stellar interiors. A set of standard tests to compare stellar hydrodynamics codes and evaluate their accuracy has not been clearly defined and organised. Although it is possible to directly characterise and compare such flows through diagnostics such as the convective turnover time, as discussed in \ct{pratt16} such flows can vary greatly in space and time, and must be observed over long times to gain meaningful statistics.

In this work we seek to find a middle ground: a set of test problems that are fundamental to stellar interiors but are also simple enough that they may be calculated quickly for the purposes of benchmarking and testing, as well as having well-defined diagnostics, to enable code comparison. We carry out this work primarily to test the accuracy of the numerical methods implemented in the MUlti-dimensional Stellar Implicit Code, \MUSIC.

\MUSIC\ is distinguished from other stellar hydrodynamics codes in that it is
both time-implicit and fully compressible. The tests collected in this work
have been chosen so that they are useful for comparing a wide variety of
physical and numerical models, including codes that are time-explicit and/or
those that implement either the anelastic or Boussinesq approximations. The
Rayleigh-Taylor, Kelvin-Helmholtz and Taylor-Green tests are relevant to a
wide range of hydrodynamical applications. The fourth test, the Hydrostatic
Equilibrium test, is specifically applied to a stellar interior, however the concept could be extended as a general test for the implementation of tabulated equations of state. Additionally the Hydrostatic Equilibrium test demonstrates for the first time the efficiency of the preconditioning technique applied in \MUSIC\ in a radiatively dominated regime.

Many astrophysical phenomena are known to exhibit dependence on non-ideal
effects, such as viscosity. In an effort to minimise non-ideal effects, codes that model such phenomena often do not contain explicit viscous terms. In such a calculation only numerical viscosity acts as a non-ideal term, entering into the solution through the truncation errors of the scheme. A particularly interesting and timely aspect of this work is the examination of the application of such codes to astrophysical phenomena. Within the context of Large Eddy Simulation (LES) calculations this approach is described as the Implicit Large Eddy Simulation (ILES) paradigm. 

The Rayleigh-Taylor and Kelvin-Helmholtz instabilities are sensitive to
non-ideal effects, which only enter into an ILES solution through truncation
errors, and vary with resolution. One might ask therefore, to what extent
should an ILES code be expected to produce solutions which converge with
increasing spatial resolution for these physical problems. For the
Rayleigh-Taylor instability we show differences in observed mixing profiles,
due to the application of two different ILES methods to a problem sensitive to
non-ideal effects. However, a systematic difference between a mixing estimate
derived under the assumption of incompressibility, and a more general estimate
was observed. For the Kelvin-Helmholtz test we show the velocity field produced in \MUSIC\ calculations exhibits the convergent properties expected from the numerical methods used. The Taylor-Green vortex has been used as a validation test for ILES codes, by monitoring the evolution of the kinetic energy. As this evolution is strongly influenced by truncation errors, we investigate the observed decay of the kinetic energy for different grid-sizes. Additionally we show how the choice of time-step can effect the decay of the kinetic energy.

This paper is structured as follows. In Section~\ref{Sec:MUSIC} we give an overview of \MUSIC. In Section~\ref{Sec:rti} we compare the mixing of a two-dimensional Rayleigh-Taylor instability produced in \MUSIC\ simulations to that produced by \ATHENA. In Section~\ref{Sec:khi} the ability of \MUSIC\ to reproduce the results for the \ct{mcnally12} Kelvin-Helmholtz instability test problem is investigated. In Section~\ref{Sec:tgv} the decay of the Taylor-Green vortex is analysed by comparing \MUSIC\ results to previous results from ILES, LES, and DNS calculations. In Section~\ref{Sec:hse} we assess the ability of \MUSIC\ to recover hydrostatic equilibrium in a radiatively dominated region of a star. We conclude in Section~\ref{Sec:conclusion}, summarising our findings and discussing their implications for future calculations of stellar interiors.
\section{The MUSIC code}
\label{Sec:MUSIC}
The \MUSIC\ code is a time-implicit, compressible hydrodynamics code. Initial development is
described in \ct{viallet11, viallet13}. Recently, \MUSIC\ has been modified to use
the Jacobian-free Newton Krylov (JFNK) method \citep{viallet16}. \MUSIC\
solves the Euler equations in the presence of external gravity and
thermal diffusion:
\begin{eqnarray}
\frac{\partial \rho}{\partial t} &=& - \vec \nabla \cdot (\rho \vec u),\label{eq:cons1}\\
\frac{\partial \rho e}{\partial t} &=& -\vec \nabla \cdot (\rho e \vec u) - p\vec \nabla \cdot \vec u + \vec \nabla \cdot (\chi \vec \nabla T),\label{eq:cons2}\\
\frac{\partial \rho \vec u}{\partial t} &=& - \vec \nabla \cdot (\rho \vec u\otimes \vec u)-\vec \nabla p + \rho \vec g\label{eq:cons3},
\end{eqnarray}
\noindent where $\rho$ is the density, $e$ the specific internal energy, $\vec
u$ the velocity, $p$ the gas pressure, $T$ the temperature, $\vec g$ the
gravitational acceleration, and $\chi$ the thermal conductivity. The gravitational acceleration does not change during a \MUSIC\ calculation. It can either be assigned a spatially constant value, or use values calculated consistently with one-dimensional model which vary with the radial coordinate. In both cases it is implemented as a body force in the momentum equation.

Boundary conditions within \MUSIC\ are implemented using ghost cells. Options include standard techniques, for example reflecting and stress-free, and less common options such as a variety of models for hydrostatic equilibrium as described in \ct{pratt16}.

Equations
\eqref{eq:cons1}-\eqref{eq:cons3} are closed by an equation of state, and an
expression for the thermal conductivity. The equation of state within
\MUSIC\ can either be taken as an ideal gas equation of state, or a tabulated
equation of state, accounting for ionisation and non-ideal effects. The thermal conductivity is given by
\begin{equation}
\label{eq:chirad}
\chi = \frac{16 \sigma T^3}{3\kappa \rho},
\end{equation}

\noindent where $\kappa$ is the Rosseland mean opacity, and $\sigma$ the Stefan-Boltzmann constant. Equation \eqref{eq:chirad} is the form of the thermal conductivity for photons. For stellar calculations the opacity is interpolated from the OPAL \citep{iglesias1996updated} and \cite{ferguson2005low} tables. 

The scalar quantities ($\rho, e$) are defined at cell centres, whereas
velocities are located at cell interfaces. To calculate advective fluxes scalar quantities and vector components are extrapolated linearly using an upstream method
\citep{van1977towards} and the reconstruction is ensured to be monotonic using the
van Leer limiter \citep{van1974towards}, resulting in a second-order total
variation diminishing (TVD) scheme.

The temporal integration is carried out using the Crank-Nicolson method
\citep{crank1947practical}, and the resulting non-linear problem is solved using
the Newton-Raphson method. At each non-linear iteration a linear problem is solved using the Generalised Minimum Residual (GMRES) method,
\citep{saad1986gmres}. A Jacobian-free Newton Krylov approach \citep[for a review see][]{knoll2004jacobian} is used to approximate the matrix-vector products required by GMRES.

The convergence of the GMRES method is improved by using a physics-based
preconditioning method, based on the work of \ct{park2009physics}. Such a preconditioner takes the form of a semi-implicit approximate solution to the full physical system. The preconditioner is semi-implicit in that it treats the stiff terms in the full system implicitly, and the remaining terms explicitly. By adjusting which terms are treated implicitly, the preconditioner can be adapted to a specific problem. Sound waves, and optionally thermal diffusion, are treated implicitly. In this work we present the first application demonstrating the efficiency of the latter case.

\subsection{Choice of Time-Step}
\label{dt}
The time-step $\Delta t$ in \MUSIC\ is adaptive and changes throughout the calculation. The time-implicit method used in \MUSIC\ allows large stable time-steps to be taken for the problems considered in this work. The practical choice of the time-step is driven by a desire for an efficient calculation, which also provides an accurate solution. \MUSIC\ will adjust $\Delta t$ in an attempt to provide a more efficient calculation. This adjustment is restricted by user-provided limits placed on the time-step. The first measure of the time-step used within \MUSIC\ is relative to the hydrodynamical CFL number:
\begin{equation}
\label{eq:cfl_hydro}
\mathrm{CFL_{hydro}} = \max\left({\frac{\left|u\right| +
    c_{\mathrm{s}}}{\Delta x}}\right) \Delta t,
\end{equation}
where $c_{\mathrm{s}}$ is sound speed, $\Delta t$ is the time-step, $\Delta x$ is the grid spacing and $u$ is the flow
velocity. A value of $\mathrm{CFL_{hydro}} = 1$ corresponds to the stability limit of a time-explicit scheme. Similarly we define the advective CFL number,
\begin{equation}
\label{eq:cfl_adv}
\mathrm{CFL_{adv}} = \max\left({\frac{\left|u\right|}{\Delta x}}\right) \Delta t.
\end{equation}
Due to the design of the physics-based preconditioner used in \MUSIC,
convergence of the linear system becomes poor for values of
$\mathrm{CFL_{adv}} > 0.5$ and as such we limit the value of this time-step
measure to be at most $0.5$ in all calculations in this work.

For calculations involving radiative effects we define the radiative CFL number:
\begin{equation}
\label{eq:cfl_rad}
\mathrm{CFL}_\mathrm{rad} = \max \left(\frac{\chi}{\Delta x^2}\right) \Delta t,
\end{equation}
with $\chi$ defined by eq. \eqref{eq:chirad}. Preliminary, low-resolution calculations can be used to determine limiting values for both $\mathrm{CFL_{rad}}$ and $\mathrm{CFL_{hydro}}$ which provide converged results in as efficient a manner as possible. In cases where multiple constraints are placed on the time-step the most restrictive one is applied.

\subsection{Passive Scalars}
\label{sec:passive_scalars}
As part of this work, \MUSIC\ has been extended to model a scalar field that is advected with the flow but does not feedback on the dynamics of the fluid. This addition, commonly referred to as ``passive scalars'' is useful for estimating the mixing and transport of physical quantities such as chemical composition and angular momentum. Example applications of passive scalars may be found in the work of \cite{madarassy2010calibrating, falkovich2005anomalous, schumacher2005very, brethouwer2005effect}. The scalars are modelled as compositions, with
density equal to the bulk fluid. The conservation equation for
a scalar $\mathrm{i}$, with concentration $\mathrm{c_i}$ is
\begin{equation}
\label{scalar_eqn}
\frac{\partial \mathrm{c_i}\rho}{\partial t} = -\nabla \cdot \left(\mathrm{c_i}\rho \vec u\right),
\end{equation}
where $\rho$ is the fluid density, and $\vec{u}$ the fluid velocity.

\MUSIC\ solves the
equation set defined by eq. \eqref{scalar_eqn} using an unpreconditioned Jacobian-free Newton Krylov
method. The same discretisation and solver settings as used for the core
solver are used for the passive scalar evolution with the exception of the
stopping criteria for the non-linear iterations. By default we do not apply
additional stopping constraints based on the passive scalar to the non-linear
iterations, instead relying on the convergence of the fluid density alone. We
take this approach to enable us to exactly reproduce results with and without
passive scalars supplementing the main equations. We assess the accuracy
of our passive scalar implementation in Section \ref{Sec:rti}.

This work shall only include cases where
two scalars are modelled. For stellar applications the number of species of
interest can take a larger value, therefore the implementation within
\MUSIC\ was designed to have no restriction on the number of passive
scalars. As eq. \eqref{scalar_eqn} evolves mass fractions, there is no
guarantee $\sum_i c_i = 1$ is maintained. For this work we apply a simple re-normalisation at the end of each time-step, but more sophisticated approaches \citep[e.g.][]{plewa1999consistent} might be required for other problems.

In the applications considered in this work the values of the scalar concentrations do not influence the evolution of the hydrodynamical state defined by equations \eqref{eq:cons1}-\eqref{eq:cons3}. For this reason we refer to the scalars as ``passive''. Eq. \eqref{scalar_eqn} may also be used to describe the evolution of chemical compositions, which do influence the core hydrodynamical state. The solution method for this situation is more complex; the scalar evolution can no longer be decoupled, and instead equations \eqref{eq:cons1}-\eqref{eq:cons3} and \eqref{scalar_eqn} must be solved as a single system.
\section{Rayleigh-Taylor instability}
\label{Sec:rti}
\subsection{Problem Description}
The Rayleigh-Taylor instability occurs when a dense fluid is accelerated, for example by gravity,
into a less dense fluid. This instability occurs in a wide range of
astrophysical applications (e.g. \cite{inogamov1999role}). The instability has also been the subject of
multiple numerical studies \citep{jun1995numerical, dimonte2004comparative}, as well as for code
validation, and comparison (e.g. \cite{liska2003comparison}).
In this test we assess the ability of \MUSIC\ to model the two-dimensional Rayleigh-Taylor
instability. We study a single mode perturbation, provided as a standard example problem\footnote{http://www.astro.princeton.edu/\textasciitilde jstone/Athena/tests}
for the \ATHENA\ code \citep{athena2005} and assess the performance of \MUSIC\ by comparison to
\ATHENA\footnote{We use \ATHENA\ version 4.2 available from https://trac.princeton.edu/Athena/wiki/AthenaDocsDownLd}. The problem is similar to that of \cite{liska2003comparison} except
in this work the domain extends to the complete wavelength of the perturbation, so that the entire mushroom is modelled. The problem is calculated
on a box defined by $-0.25<x<0.25$ and $-0.75<y<0.75$. The aspect ratio of the box is chosen so that the primary instability remains far from the boundaries for the times considered. A constant
gravitational acceleration of magnitude $g=0.1$ acts in the negative
y-direction. The density is given by,
\begin{equation}
\rho = \begin{cases} 2.0 &\mbox{if } y > 0.0, \\
1.0 &\mbox{if } y \le 0.0\end{cases}.
\end{equation}
The pressure is calculated by solving the equation of hydrostatic equilibrium,
and is given by
\begin{equation}
P = P_0 - \rho\,g\,y.
\end{equation}
where $P_0 = 2.5$. The equation of state is an ideal-gas law, with $\gamma = 1.4$. The Rayleigh-Taylor instability is sensitive to choices of the initial
perturbation \citep{ramaprabhu2005numerical}. The instability may be seeded by either perturbing the
interface, or the velocity. In this work the instability is seeded through the velocity. The
velocity perturbation is given by\footnote{Equation \eqref{rti_pert} is taken from the \ATHENA\ source code.}
\begin{equation}
\label{rti_pert}
v_y = 0.0025\left[1+\cos\left(4\pi x\right)\right]\left[1+\cos\left(\frac{4}{3}
  \pi y\right)\right].
\end{equation}
We use dimensionless units, but we note the pressure scale height varies between approximately 25.752 at the bottom of the domain and approximately 11.752 at the top. The linear growth rate of the Rayleigh-Taylor instability depends on the gravitational acceleration, and the dimensionless Atwood number which takes a value of $1/3$ in this case. To compare to more realistic values for stellar cases, the scaling implied by the pressure scale height ($x_0$) and the gravitational acceleration ($g_0$) when combined with a scaling for density ($\rho_0$) provide a normalisation for the Euler equations in the presence of external gravitational acceleration, which is the system being described by this test case. Boundary conditions in the vertical directions are calculated by linearly extrapolating the temperature. The density is then calculated according
to the equation of hydrostatic equilibrium. Reflective and stress-free boundary conditions are
applied to the velocity components. Periodic boundary conditions are applied
in the horizontal directions.
\begin{figure}[t]
  \centering
  \includegraphics[width=0.8\linewidth]{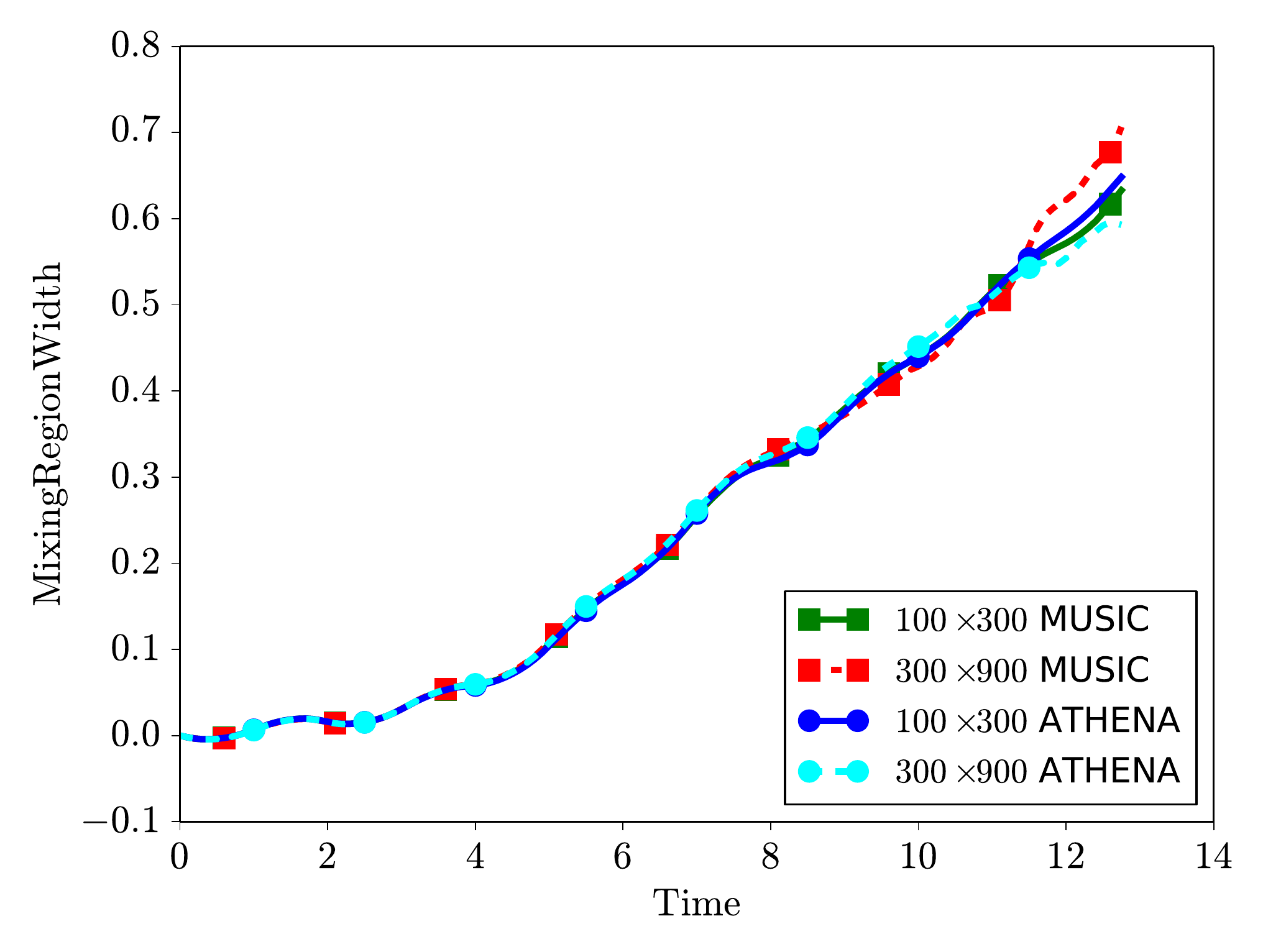}
  \caption{Evolution of mixing region width using the method of Cabot and Cook (2006).}
  \label{fig:RTI_mixing_rho}
\end{figure}
The evolution of the Rayleigh-Taylor instability, particularly in the non-linear phase is strongly sensitive to non-ideal effects (e.g. viscosity, \citet{cabot2006reynolds,lim2010nonideal}).
\MUSIC\ includes no explicit viscosity, and thus non-ideal effects enter only through errors introduced by the numerical scheme. To aid comparison we run \ATHENA\ without including explicit viscosity. Late time evolution is influenced by secondary Kelvin-Helmholtz instabilities. The break up of the interface between the two fluids through secondary instabilities is strongly dependent on the numerical scheme as discussed by \cite{liska2003comparison}. Comparisons between two codes run without explicit viscosity must be performed with care, because the growth rate of the primary instability, and the development of secondary instabilities are both sensitive to the non-ideal effects caused by the truncation errors of the different numerical schemes.
\subsection{Mixing Region Width Calculation}
We quantify the mixing in a given simulation by calculating a mixing region width. By measuring the integrated amount of mixing in a given region such a diagnostic should provide insight into the effect the Rayleigh-Taylor instability could have on a more complex physical system. Mixing of different chemical species through the Rayleigh-Taylor instability could influence stellar structure, convective stability, and nuclear burning rates by altering composition. A mixing width measures the extent to which the two initially separate species have been mixed. This width is calculated by integrating the horizontally averaged mixing fraction, which we use two methods to calculate. The first is the method of \cite{cabot2006reynolds}, developed in the incompressible limit. The second method is to use passive scalars, which capture compressible effects. As discussed in \cite{miczek2015new, guillard1999behaviour}, low Mach number flows, which typically occur in stellar interiors, approach the incompressible limit. The Rayleigh-Taylor instability is a subsonic phenomenon. For a grid size of $100 \times 300$ we obtain a maximum Mach number of $0.2607$ with \MUSIC, and consequently compressible effects are expected to be small. A comparison between the two methods of estimating mixing should provide insight into the role compressible effects play in mixing in the Rayleigh-Taylor instability.

Following \cite{cabot2006reynolds}, the fraction of dense material in a cell,
$X_{\mathrm{H}}$, is
\begin{equation}
\label{eq:rho_mix}
X_{\mathrm{H}} = \frac{\rho - \rho_{\mathrm{L}}}{\rho_{\mathrm{H}} -
  \rho_{\mathrm{L}}},
\end{equation}
where $\rho$ is the (volume averaged) density of a computational cell, $\rho_{\mathrm{H}}$
is the initial density of the heavy fluid (2.0 in this work), and $\rho_{\mathrm{L}}$
is the density of the light fluid (1.0). The fraction of mixed fluid is
\begin{equation}
X_{\mathrm{M}} = \begin{cases} 2X_{\mathrm{H}} &\mbox{if } X_{\mathrm{H}} \le 0.5 \\
2\left(1-X_{\mathrm{H}}\right) &\mbox{if } X_{\mathrm{H}} > 0.5\end{cases}.
\end{equation}
The mixing region width is then defined as
\begin{equation}
\label{mixing_width}
h = \int_{-\infty}^{+\infty}X_{\mathrm{M}}\left(\left<X_{\mathrm{H}}\right>\right)dy,
\end{equation}
where $\left<X_{\mathrm{H}}\right>$ is the average fraction of dense material
in a horizontal layer.
The Rayleigh-Taylor instability is also analysed using two passive scalar fields,
each evolved according to eq. \eqref{scalar_eqn}. One passive scalar marks the
dense fluid, the other marks the lighter fluid,
\begin{equation}
\left(c_1,c_2\right) = \begin{cases} \left(1.0,0.0\right) &\mbox{if } y \le 0.0 \\
\left(0.0,1.0\right) &\mbox{if } y > 0.0 \end{cases}.
\end{equation}
Passive scalars allow the calculation of the mixing region width defined by eq. \eqref{mixing_width} without the assumption of incompressibility. In this case the mixing fraction is,
\begin{equation}
X_{\mathrm{M}} = 2.0 \min\left(c_1,c_2\right).
\end{equation}
Having calculated the mixing fraction, the mixing width can once again be calculated using eq. \eqref{mixing_width}.
\begin{figure*}[t]
  \centering
  \centering\includegraphics[width=0.8\linewidth, trim= 0cm 0 0 0,clip=true]{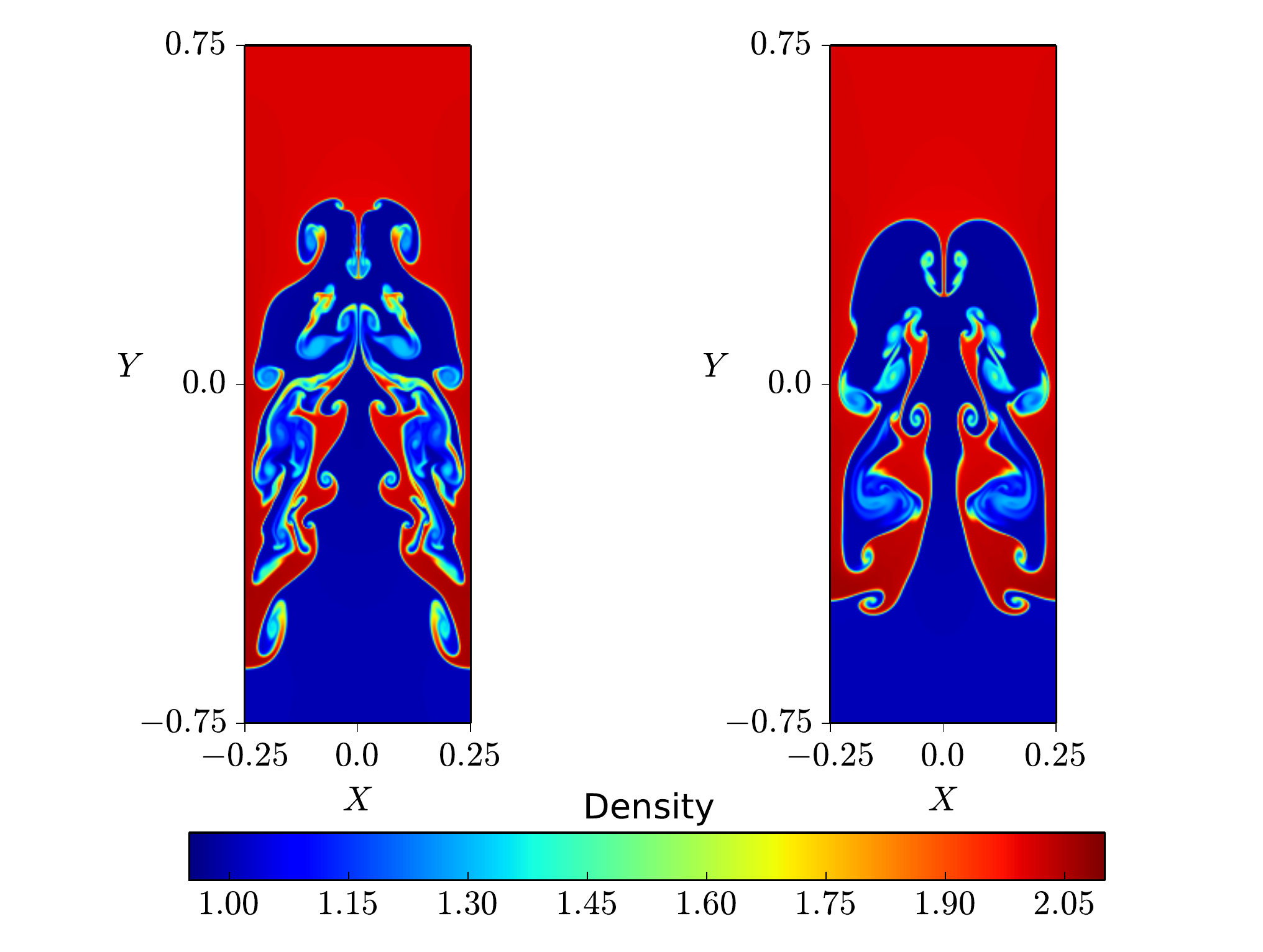}
  \caption{Final density plots for identical Rayleigh-Taylor calculations performed with (left) \MUSIC\ and (right) \ATHENA\ each with a grid size of $\left(300,900\right)$.}
  \label{fig:RTI_density}
\end{figure*}

As \MUSIC\ is a time-implicit code, the time-step is not restricted by the CFL condition. However, concerns
over accuracy and efficiency may provide practical limitations. Given the sensitivity of the Rayleigh-Taylor instability, and to simplify comparison to \ATHENA\ we carry out \MUSIC\ calculations with a fixed value of $\mathrm{CFL_{Hydro}} = 0.8$, which is the default value provided for the \ATHENA\ example. This choice does not take advantage of the large time-step allowed by the time-implicit method implemented within \MUSIC, it is chosen to simplify comparison with the \ATHENA\ code.

\subsection{Effect of Grid Size}
We now study the effect of grid size on the evolution of the
the Rayleigh-Taylor instability. At early times,
the evolution is expected to be dominated by the initial perturbation. Any
differences in observed mixing region widths should be attributed to failure
to resolve the initial perturbation or changes in non-ideal effects caused by varying the grid size. At later times, secondary instabilities
can become more important. \cite{liska2003comparison} show that less dissipative codes experience a higher rate of secondary instability, and a resulting break-up of the fluid interface. 

For this test we use the un-preconditioned JFNK time-integration method in
\MUSIC\ to compare with results from the \ATHENA\ code. We carry out identical calculations using two different two-dimensional grids. Grid sizes of $100\times300$ and $300\times900$ ensure the aspect ratio of the computational cells is equal to $1.0$. As the effective viscosity of an ILES calculation depends on the truncation errors of the scheme, differences between results from different codes at a specific grid size should be expected. However, as both \MUSIC\ and \ATHENA\ are spatially second order codes each should experience similar behaviour with increasing grid size. At higher resolution, because non-ideal effects become less significant, secondary instabilities should become more prevalent. The emergence and evolution of secondary instabilities are not seeded by the initial conditions, but through the truncation errors of a given scheme. Therefore as the secondary instabilities grow differences between different schemes may increase.

The evolution of the mixing region width, for \MUSIC\ and \ATHENA\ using the method of \cite{cabot2006reynolds} is shown in Fig.~\ref{fig:RTI_mixing_rho}. At early times the \cite{cabot2006reynolds} mixing region width takes an un-physical negative value in both \MUSIC\ and \ATHENA\ calculations. This is due to the effects of compressibility not being taken into account in this definition of the mixing region width. The high and low grid size calculations with \MUSIC\ diverge around $t=12.0$, whereas the two \ATHENA\ calculations show more similar values. That \MUSIC\ results show a stronger dependence on grid size at later times may be indicative of secondary instabilities playing a stronger role in the evolution of the mixing. The influence of secondary instabilities may be enhanced through the non-exact time integration within \MUSIC.

Fig.~\ref{fig:RTI_density} shows that calculations from both \ATHENA\ and \MUSIC\ perfectly maintain the symmetry present in the initial conditions. This result demonstrates a physically important feature of the GMRES algorithm: if the matrix vector products respect a given physical symmetry, GMRES is able to produce an approximate solution to the linear problem which is also symmetric. The JFNK method approximates matrix-vector products through the evaluation of the non-linear residual of the full system. In order to obtain a fully symmetric solution, the matrix-vector products must be exactly symmetric. Given the non-associativity of floating point arithmetic, care must be taken in the order of calculations\footnote{We note that, in this respect codes written in C (or C++), such as \ATHENA, have an advantage over codes written in Fortran. The C and C++ standards dictate compilers must respect the order of calculations, whereas Fortran codes are only restricted by order implied by parentheses. Furthermore, not all Fortran compilers (e.g. Intel) follow this restriction by default. For example, see the discussion in \cite{corden2009consistency}.}.

It is also evident in Fig.~\ref{fig:RTI_density} that the mixing in the \MUSIC\ calculation becomes asymmetric in the vertical direction; the dense fluid penetrates further into the lighter fluid than the lighter fluid does into the dense fluid. The \ATHENA\ calculation remains more symmetric in this respect. Such enhanced mixing in the lower domain may be caused by the enhanced secondary instabilities discussed previously.

In contrast to the agreement between the codes in the calculation of the mixing region width Fig.~\ref{fig:RTI_density} shows significant differences in the development of secondary instabilities. These differences may be caused by differences in the initial conditions, or by differences in the numerical technique applied. The perturbation specified by eq. \eqref{rti_pert} is identical in a continuous sense, but the exact discrete form will differ between \MUSIC\ and \ATHENA. \ATHENA\ uses co-located variables, whereas \MUSIC\ applies a staggered grid approach. The secondary instabilities which dominate the differences between the two codes are not seeded explicitly by the initial perturbation, and enter into the initial conditions only through discretisation errors. Furthermore truncation errors can seed and enhance secondary instabilities during the course of a simulation. In particular differences between the spatial reconstruction methods used by \MUSIC\ and \ATHENA\ will compound differences between the two results. That \MUSIC\ and \ATHENA\ obtain similar mixing region widths despite these differences it should be concluded that for the times considered the primary instability dominates mixing.

In addition to verifying the core hydrodynamic method with MUSIC we also use the Rayleigh-Taylor instability to test the implementation of the passive scalars discussed within Subsection~\ref{sec:passive_scalars}. We investigate the impact of not enforcing the non-linear convergence of the passive scalars. To this end we compare two calculations: firstly a calculation where the passive scalars are not accounted for in the non-linear convergence, and secondly a calculation where we require the corrections to the passive scalars to converge to the same level of accuracy as the primary variables.
\begin{figure}[t]
  \centering
  \includegraphics[width=0.8\linewidth]{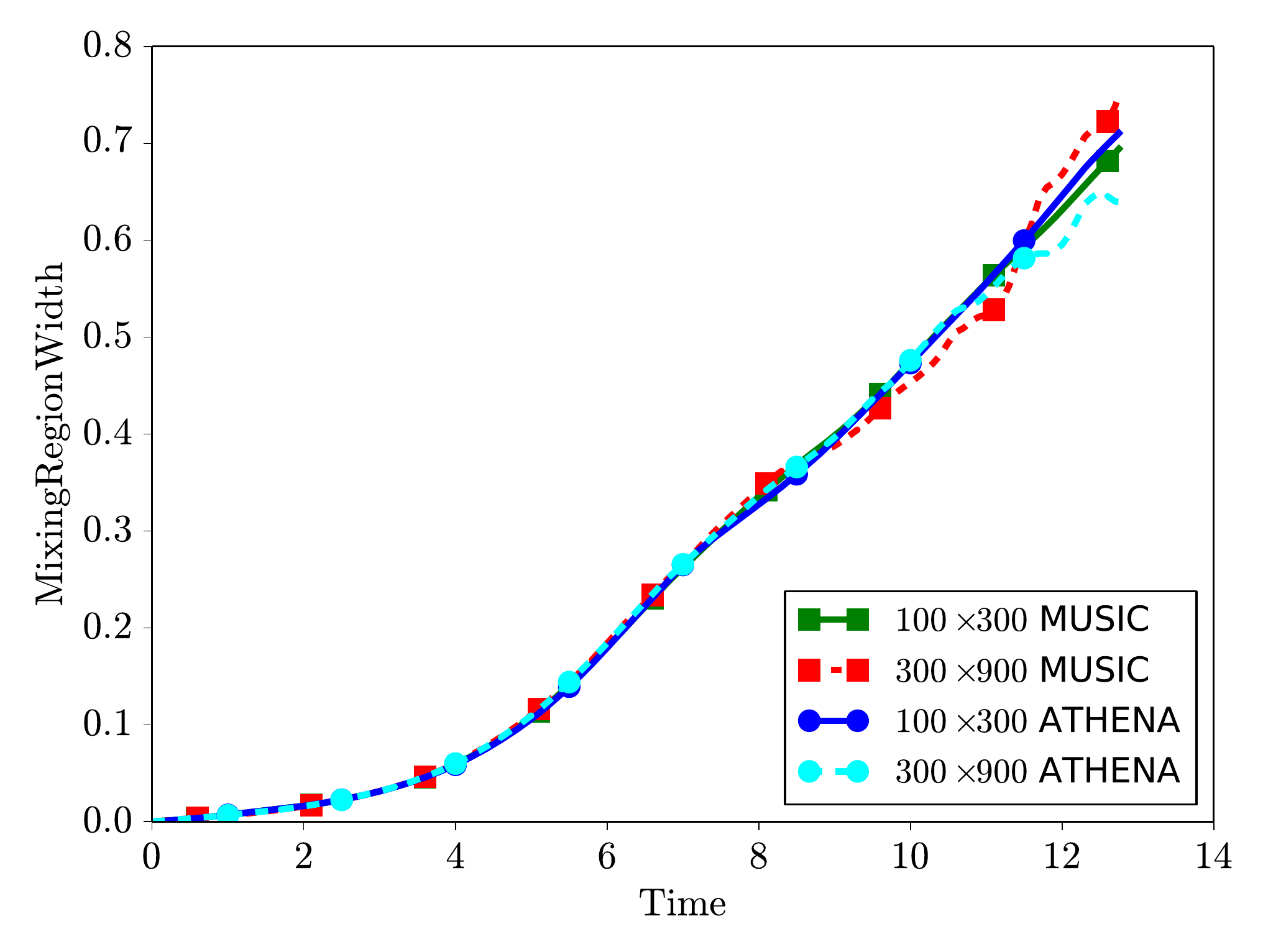}
  \caption{Evolution of mixing region width using passive scalars.}
  \label{fig:RTI_mixing_scalar}
\end{figure}
We compare differences in the measured mixing region width calculated using the volume fractions of the passive scalars. No assumption of incompressibility is made; the passive scalars act as a dye to measure the amount of mixing within each grid cell. We do not observe any significant changes in the mixing region width between the two calculations, but we do note that by enforcing non-linear convergence of the passive scalars the run-time increases by approximately 10\%. In all further calculations we do not enforce the convergence of the passive scalars, but we stress that such an approach should be assessed for a given application.

\ATHENA\ can also, optionally, evolve passive scalars. We now compare results obtained using passive scalars in \MUSIC, to those obtained with \ATHENA. The passive scalars evolved by \MUSIC\ and \ATHENA\ also maintain the symmetry of the solution exactly. Fig.~\ref{fig:RTI_mixing_scalar} compares the mixing region widths calculated using passive scalars. In all cases the mixing region width calculated using passive scalars is larger than that observed using the fluid density, suggesting that the assumption of incompressibility systematically underestimates mixing in the case considered here. The un-physical, early time negative mixing region width observed with the \cite{cabot2006reynolds} method is not observed in the scalar measurements. For both sets of calculations the mixing region width calculated using passive scalars is larger than that using the method of \cite{cabot2006reynolds} indicating that the assumption of incompressibility systematically underestimates mixing in this case. Furthermore we can conclude that compressible effects are comparable in the calculations of \MUSIC\ and \ATHENA.
\begin{figure*}
  \centering
  \centering\includegraphics[width=1.1\linewidth, trim= 2cm 0 0 0,clip=true]{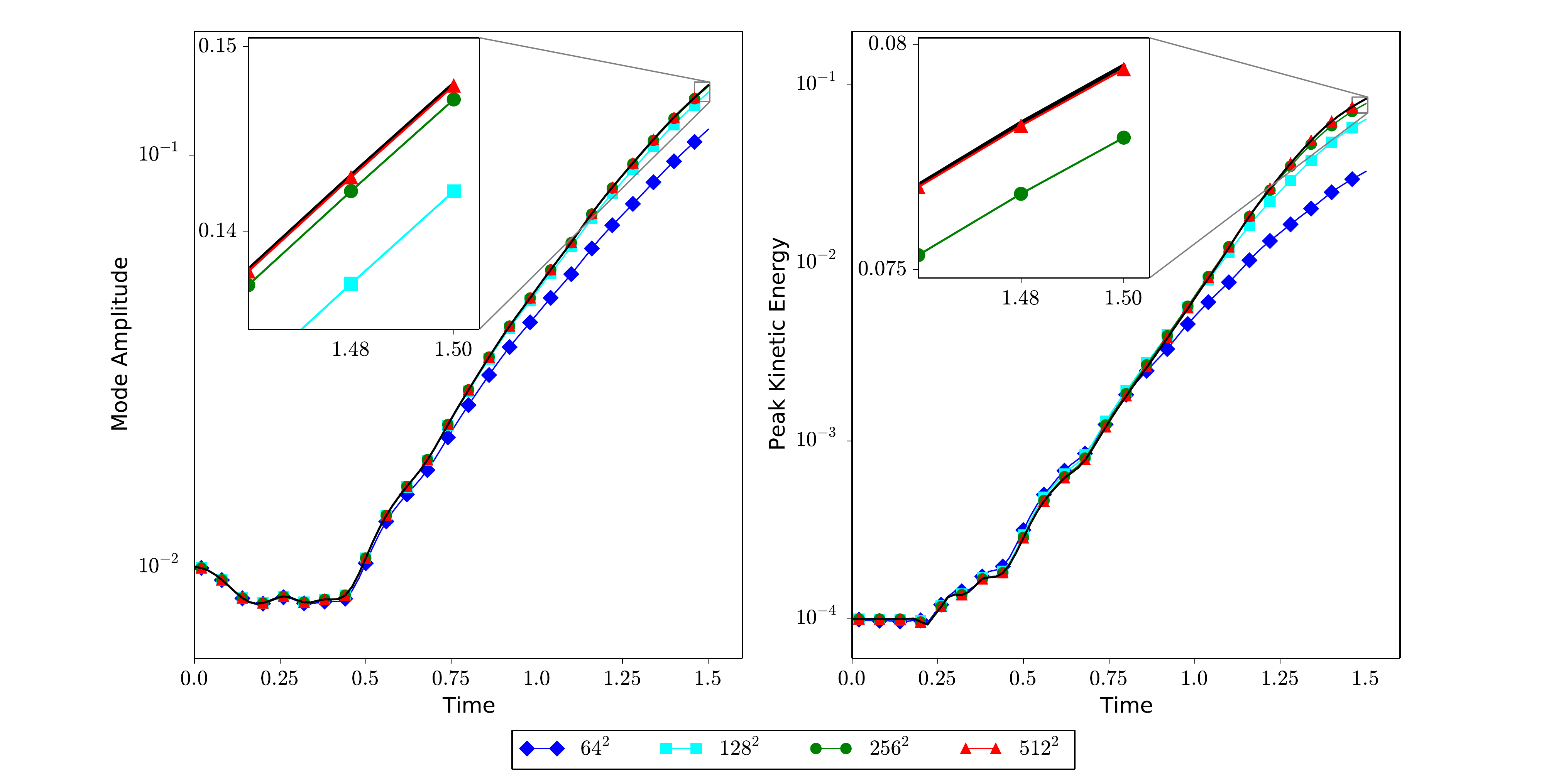}
  \caption{Evolution of the mode amplitude and peak vertical kinetic energy for the Kelvin-Helmholtz instability. The \ct{mcnally12} solution is shown as a black line. Units are dimensionless.}
  \label{fig:KHI_split}
\end{figure*}

Two methods of estimating Rayleigh-Taylor induced mixing have been compared using the \MUSIC\ and \ATHENA\ codes. Differences are expected because both codes were used as ILES codes. The \MUSIC\ code appears more sensitive to secondary instabilities. Despite this stronger sensitivity, in both cases a systematic under estimate of mixing is seen when using the method of \cite{cabot2006reynolds} under the assumption of incompressibility.
\section{Kelvin-Helmholtz instability}
\label{Sec:khi}
\subsection{Problem Description}
The Kelvin-Helmholtz instability has been invoked to explain mixing in novae explosions \citep{casanova2011kelvin} as well as vertical mixing in stellar interiors due to differential rotation \citep{bruggen2001mixing}. Test problems for the instability exist in many forms \citep{wang2010combined,agertz2007fundamental}, here we investigate the test case presented by \ct{mcnally12}.

\ct{mcnally12} present a set of initial conditions that does not contain sharp
discontinuities. Additionally a reference solution, calculated using the
\PENCIL\ code
\citep{brandenburg2001hydromagnetic,lyra2008global}\footnote{Available from
  http://pencil-code.nordita.org/} was provided, in terms of a peak kinetic
energy, and the mode amplitude, or a resolution of $4096^2$. The uncertainty in the solution provided was calculated using Richardson extrapolation \citep{roache1998verification,roache1994perspective}.

We calculate the mode amplitude and peak kinetic energy for a series of \MUSIC\ simulations that use different grid sizes (Fig.~\ref{fig:KHI_split}). The peak kinetic energy appears to match the solution of  \ct{mcnally12} for grid sizes greater than $512^2$ while the value obtained for the mode amplitude shows good correspondence with the reference solution for grid sizes greater than $256^2$. \ct{mcnally12} calculated both the peak kinetic energy and the mode amplitude using a selection of grid based and meshless codes. For the grid based codes, \ct{mcnally12} also showed a smaller error for the mode amplitude compared to that of the peak kinetic energy, at a given grid size.

As in the case of the Rayleigh-Taylor instability, the Kelvin-Helmholtz instability is sensitive to non-ideal effects. In a recent work \ct{lecoanet2016validated} considered the possibility of defining an effective Reynolds number for Kelvin-Helmholtz instabilities calculated with differing grid size in the ILES framework. The attribution of an effective Reynolds number was successful for cases without a density contrast. Using the \ATHENA\ code, \ct{lecoanet2016validated} were able to find a good match between cases with and without explicit viscosity and attribute this to an approximate increase in Reynolds number with an increase in grid size. The comparison was inconclusive for cases with a density contrast such as the case considered here. For all cases, an increase in grid size corresponds to a decrease in non-ideal effects, but defining an effective Reynolds number is problem dependent, and is not always possible. Care should be taken in interpreting the convergence of such simulations.
\subsection{Effect of Grid Size}
We consider the convergence of the velocity field for the Kelvin-Helmholtz instability. We study the behaviour of the vertical velocity component with grid size. The spatial discretisation in \MUSIC\ varies between first and second order, due to the application of a gradient limiter. Therefore, for a problem dominated by discontinuities, the convergence of the scheme should be first order, whereas for a smooth solution one should expect the solution to converge at second order. 

In the absence of an analytic solution for the velocity field we calculate errors with respect to the highest grid size solution we obtain, $4096^2$. This does favourably bias the solution produced by MUSIC, in effect it will mask any systematic error in the solution. We rule out the presence of a systematic error, based on the ability of \MUSIC\ to reproduce the peak kinetic energy and the mode amplitude provided by \ct{mcnally12}, and reproduced by several codes in the same study. In the absence of a systematic error, such a study provides an insight and measure of how the error reduces with increasing grid size.
\begin{figure}
  \centering
  \includegraphics[width=0.8\linewidth, trim= 0 0 0 0]{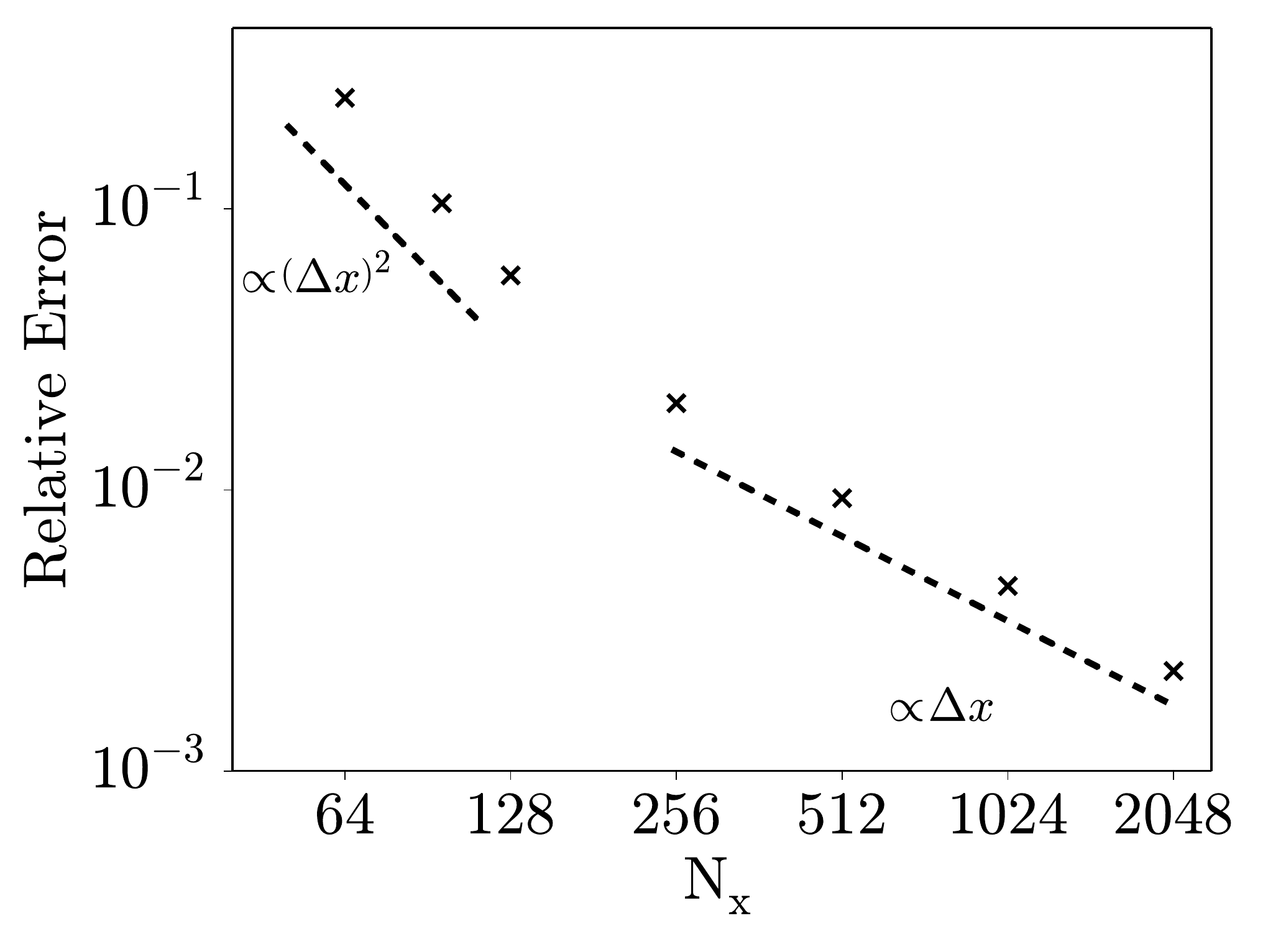}
  \caption{Relative error, as defined by eq. \eqref{eq:rel_error}, in the vertical velocity component for the Kelvin-Helmholtz test for different numbers of grid points in the x-direction. In all calculations $\mathrm{N_y} = \mathrm{N_x}$. Dashed lines indicate regions where second and first order convergence is observed.}
  \label{fig:KHI_error}
\end{figure}
In order to calculate the relative error at each grid size we coarsen the high grid size solution to the lower grid size using a volume averaging approach, following the method of \ct{toth2000b}. Such an approach is consistent with the finite volume formulation of MUSIC. Having coarsened the high grid size data we calculate the relative absolute value defined as,
\begin{equation}
\label{eq:rel_error}
\epsilon = \frac{\sum \left|v_y^{\mathrm{low}} - \overline{v_y^{\mathrm{high}}}\right|}{\sum \left|\overline{v_y^{\mathrm{high}}}\right|},
\end{equation}
where $v_y^{\mathrm{low}}$ is the low grid size data, and $\overline{v_y^{\mathrm{high}}}$ is the coarsened high grid size data. Summations are carried out over all grid cells. We plot the variation of this error with grid size in Fig.~\ref{fig:KHI_error}.
\begin{figure*}
   \centering
   \parbox{0.45\linewidth}{\centering \includegraphics[width=0.8\linewidth, trim= 60 0 60 20]{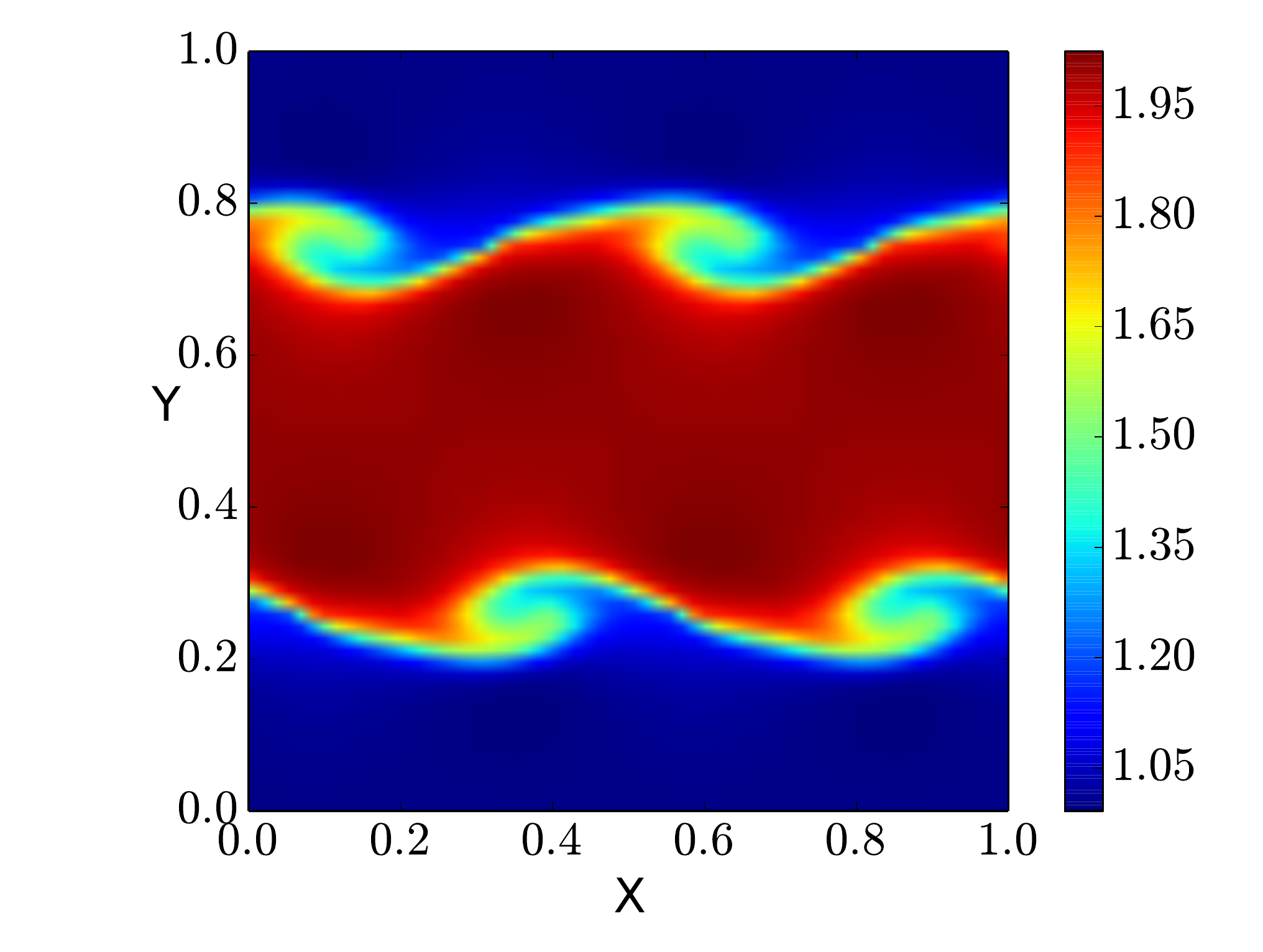}}
   \parbox{0.45\linewidth}{\centering \includegraphics[width=0.8\linewidth, trim= 60 0 60 20]{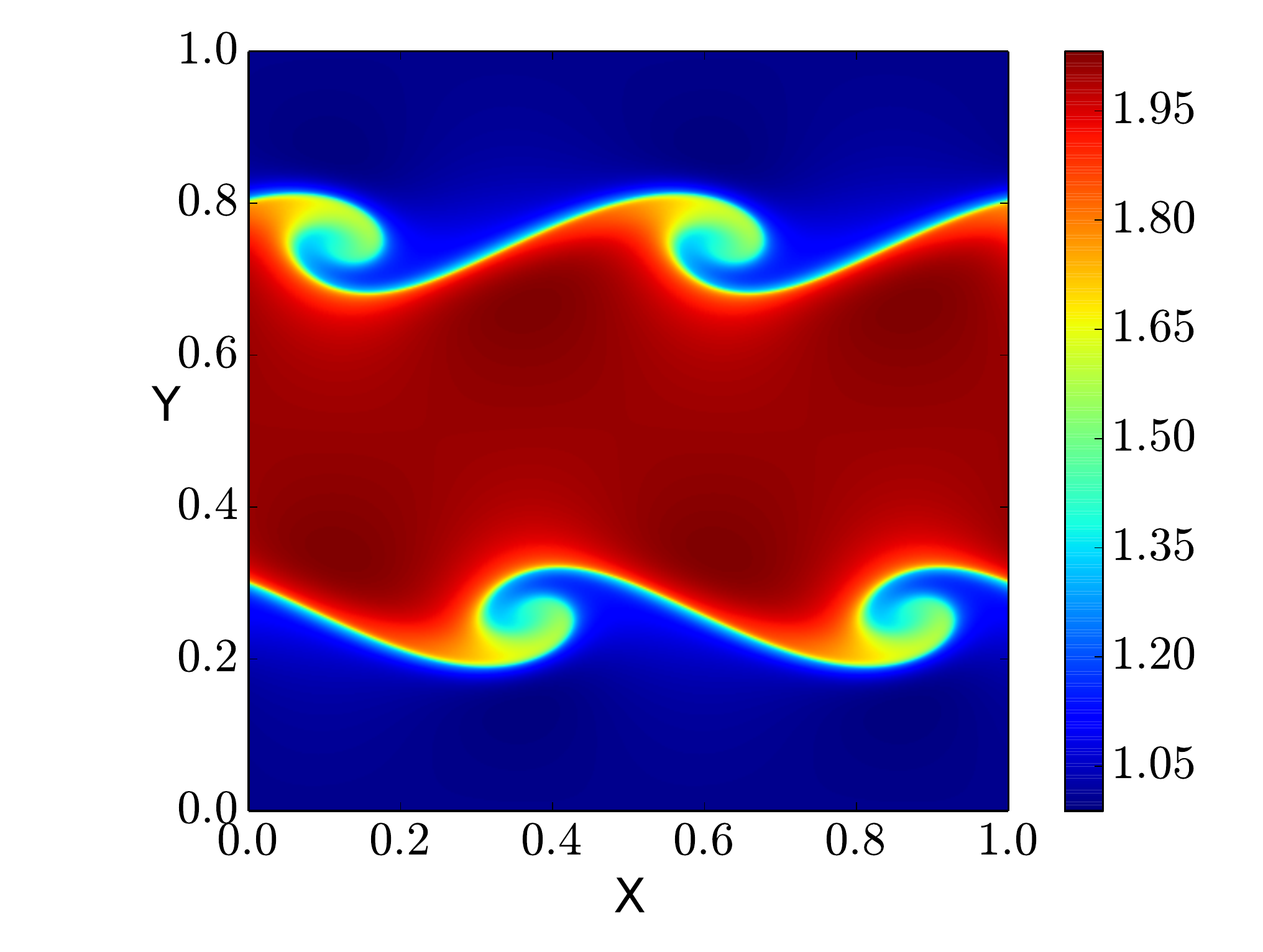}}
   \caption{Visualisation of the density in the Kelvin-Helmholtz problem for (left) grid size $64^2$ and (right) grid size $2048^2$ at $t=1.5$.}
   \label{fig:KHI_rho}
\end{figure*}
At low grid sizes the error converges with approximately second order with respect to the grid spacing as expected. As grid size increases (beyond $256^2$) the convergence tends towards first order. This indicates that the error is dominated by regions in which the solution is discontinuous, causing the spatial scheme to switch to first order. The density at $t=1.5$ is shown in Fig.~\ref{fig:KHI_rho} for grid sizes of $64^2$ and $2048^2$. In the $64^2$ case the interface between the layers of different density is smeared across several grid cells, whereas it remains sharper in the $2048^2$ case. By comparing Fig.~\ref{fig:KHI_rho} and Fig.~\ref{fig:KHI_error_location} it is clear the error is concentrated around the region of the density jump. Such a localisation in error was also shown in Figs. 5 and 6 of \ct{mcnally12}.
\begin{figure*}
   \centering
   \parbox{0.45\linewidth}{\centering \includegraphics[width=0.8\linewidth, trim= 60 0 60 20]{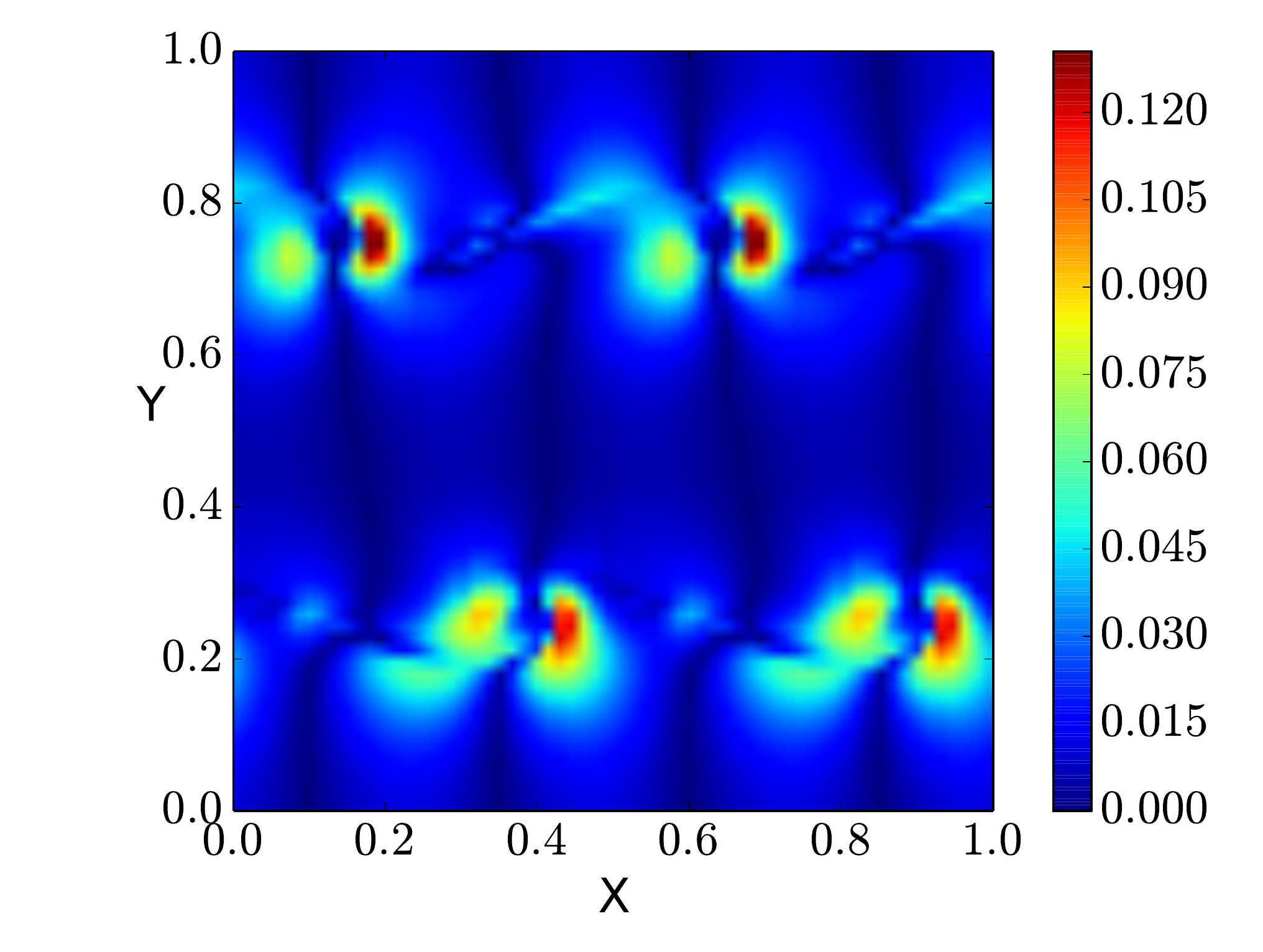}}
   \parbox{0.45\linewidth}{\centering \includegraphics[width=0.8\linewidth, trim= 60 0 60 20]{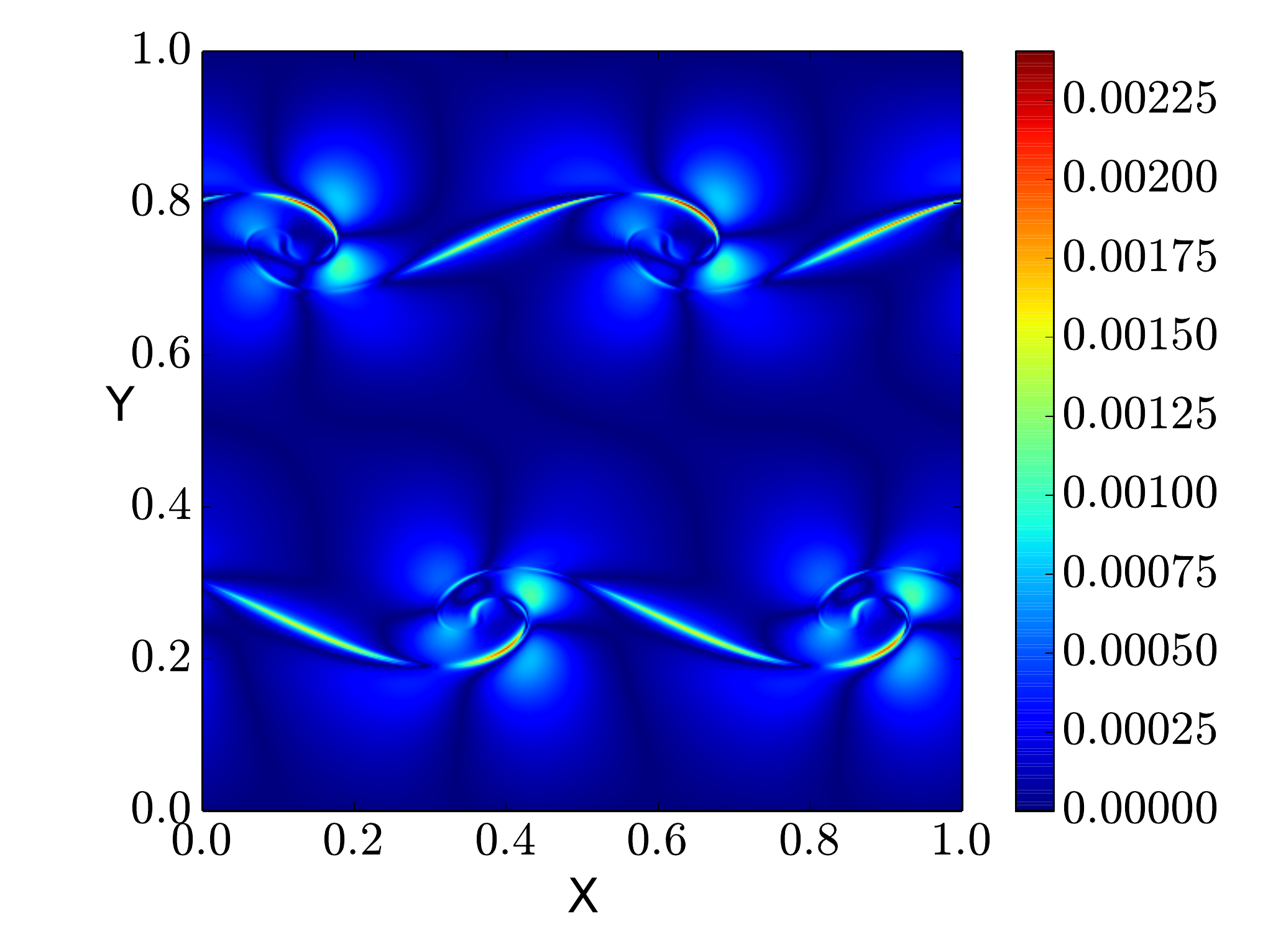}}
   \caption{Visualisation of the relative errors for the Kelvin-Helmholtz test compared to the highest grid size $4096^2$ for (left) grid size $64^2$ and (right) grid size $2048^2$ at $t=1.5$.}
   \label{fig:KHI_error_location}
\end{figure*}

We have demonstrated the ability of the MUSIC code to reproduce key diagnostics of the Kelvin-Helmholtz instability compared with those reported by \ct{mcnally12}. Although MUSIC does not include explicit viscous terms, we have also demonstrated a reduction in error for the velocity field consistent with the numerical methods applied. 
\section{Decay of the Taylor-Green vortex}
\label{Sec:tgv}
\subsection{Problem Description}
The decay of the Taylor-Green vortex \citep{taylorgreen} has been used as a benchmark for the
modelling of turbulent decay in multiple studies. We follow the
study of \cite{drikakis2007simulation} (from here on referred to as ``DFGY2007'') which assesses the ability of the Monotone Implicit
Large Eddy Simulation (MILES) method to reproduce features of vortex decay
observed when studying the problem with conventional Large Eddy Simulation (LES)
and Direct Numerical Simulations (DNS).
We do not explicitly attempt to assess the validity of the MILES
paradigm, as numerous works on this topic already exist, we simply
compare \MUSIC\ to established MILES calculations. This provides an
opportunity to investigate the ability of the spatial discretisation in \MUSIC\
to perform as a MILES code. We assess any possible side-effects
the time-implicit method has on MILES calculations.

The initial conditions of the Taylor-Green vortex are given by
\begin{align}
u_x(x,y,z) &= u_0 \sin x \cos y \cos z,\\
u_y(x,y,z) &= - u_0 \cos x \sin y \cos z,\\
u_z(x,y,z) & = 0.
\end{align}
The domain has an arbitrary uniform density of $\rho_0=1.0$. The initial pressure field is
\begin{align}
p(x,y,z) = p_0 + \frac{1}{16} \rho_0 u_0^2 (2 + \cos 2z)(\cos 2x + \cos 2y).
\end{align}
The domain is a cube with edge lengths of $2\pi$, and boundary conditions are periodic in all directions. As in DFGY2007 dimensionless units are used.

In a previous study with \MUSIC\ \citep{viallet16} $u_0$ was fixed to 1.0, and $p_0$ was adjusted to simulate the decay of the Taylor-Green vortex for a range of Mach numbers, $10^{-1} \leq M_s \leq 10^{-6}$. However in this work, we adjust $p_0$ so that the initial peak Mach number is $M_s = 0.28$, as in DFGY2007. Therefore, in addition to verifying \MUSIC\ through comparison to a range of ILES, conventional LES, and DNS simulations, we can also investigate possible compressive effects through comparison to \cite{viallet16}.
\subsection{Effect of Time Step}
\begin{figure}
   \centering
   \includegraphics[width=0.8\linewidth, trim= 60 0 60 20]{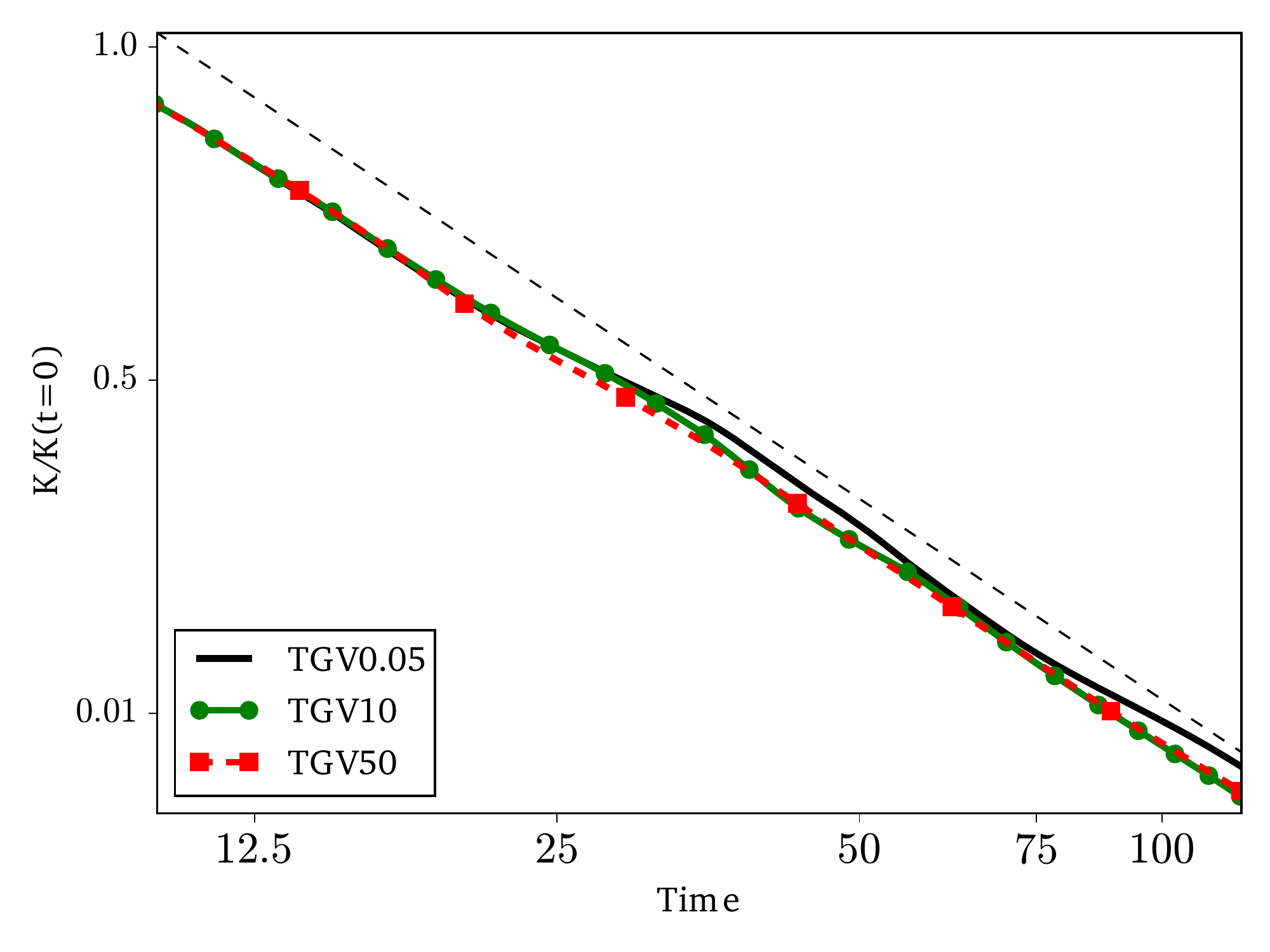}
   \caption{Evolution of the volume averaged, normalised kinetic energy in Taylor-Green vortex simulations. The three simulations shown are identical except for the limitation on the time-step based on the hydrodynamic CFL number. The black dashed line shows a $t^{-2}$ decay.}
   \label{fig:TGV_timestep}
\end{figure}

Within ILES calculations the dissipation of kinetic energy occurs through the truncation errors of the scheme. We first investigate the ability of \MUSIC\ to reproduce kinetic energy evolution for different limits on the adaptive time-step. We carry out three calculations, at a grid size of $256^3$. In the first calculation we fix the $\mathrm{CFL}_{\mathrm{hydro}} = 0.05$ (``TGV0.05''). In the remaining two runs, we impose limits on the hydrodynamical CFL number, defined in eq. \eqref{eq:cfl_hydro}, limiting $\mathrm{CFL}_{\mathrm{hydro}} \leq 10$ (``TGV10''), and $\mathrm{CFL}_{\mathrm{hydro}} \leq 50$ (``TGV50''). We show the evolution of the kinetic energy (normalised to its initial value) in Fig.~\ref{fig:TGV_timestep}. For early times the kinetic energy for all three simulations is very similar. 

A decay of $t^{-1.2}$ of the kinetic energy is predicted by Saffman's law \citep{saffman1967note} for homogeneous high Reynolds number turbulence. \cite{skrbek2000decay} interpret decays faster than $t^{-1.2}$ as being caused by viscous corrections to the high Reynolds number result. At later times the finite-size of the domain results in a quadratic decay of kinetic energy , as discussed by \cite{lesieur20003d}. Such a decay has also been observed experimentally by \cite{stalp1999decay}. 

We fit power-law decays for the kinetic energy in \MUSIC\ calculations for two time periods. The first spans to $8.4\leq t \leq 10$. This covers the time from the peak dissipation rate shown in Fig~\ref{fig:TGV_RES}, and the point at which the decay takes on a steady, steeper decay. Power-law decays are also fit for the period $t>10$. Both sets of values are recorded in Table \ref{table:tgv_table}. For the fits to the early ($8.4\leq t \leq 10$) time we find a values between the high and low Reynolds number predictions from Saffman's law, indicating the calculations are in neither extreme regime.

All calculations show similar evolution, up until $t=20$, at which point the calculation with the least restrictive time-step constraints (TGV50) shows a slightly increased rate of dissipation. The TGV10 case matches the fixed time-step calculation until approximately $t=25$ at which point it too shows a slight increase in dissipation rate when compared to the fixed CFL number calculation. At later times the TGV50 and TGV10 calculations show similar kinetic energy, both slightly less than the fixed CFL number calculation. All three data sets show decays slightly slower than $t^{-2}$. These results can be compared to Fig. 5 of DFGY2007. This shows four ILES and three LES schemes producing an approximate decay of kinetic energy as $t^{-2}$. All schemes shown in DFGY2007 show fluctuations around the $t^{-2}$ decay, indeed the differences seen in the three sets of calculations using \MUSIC\ appear smaller than those observed between different ILES schemes in DFGY2007. 
\subsection{Effect of Grid Size}
\begin{figure}
  \centering
  \includegraphics[width=0.8\linewidth, trim= 0 0 0 0]{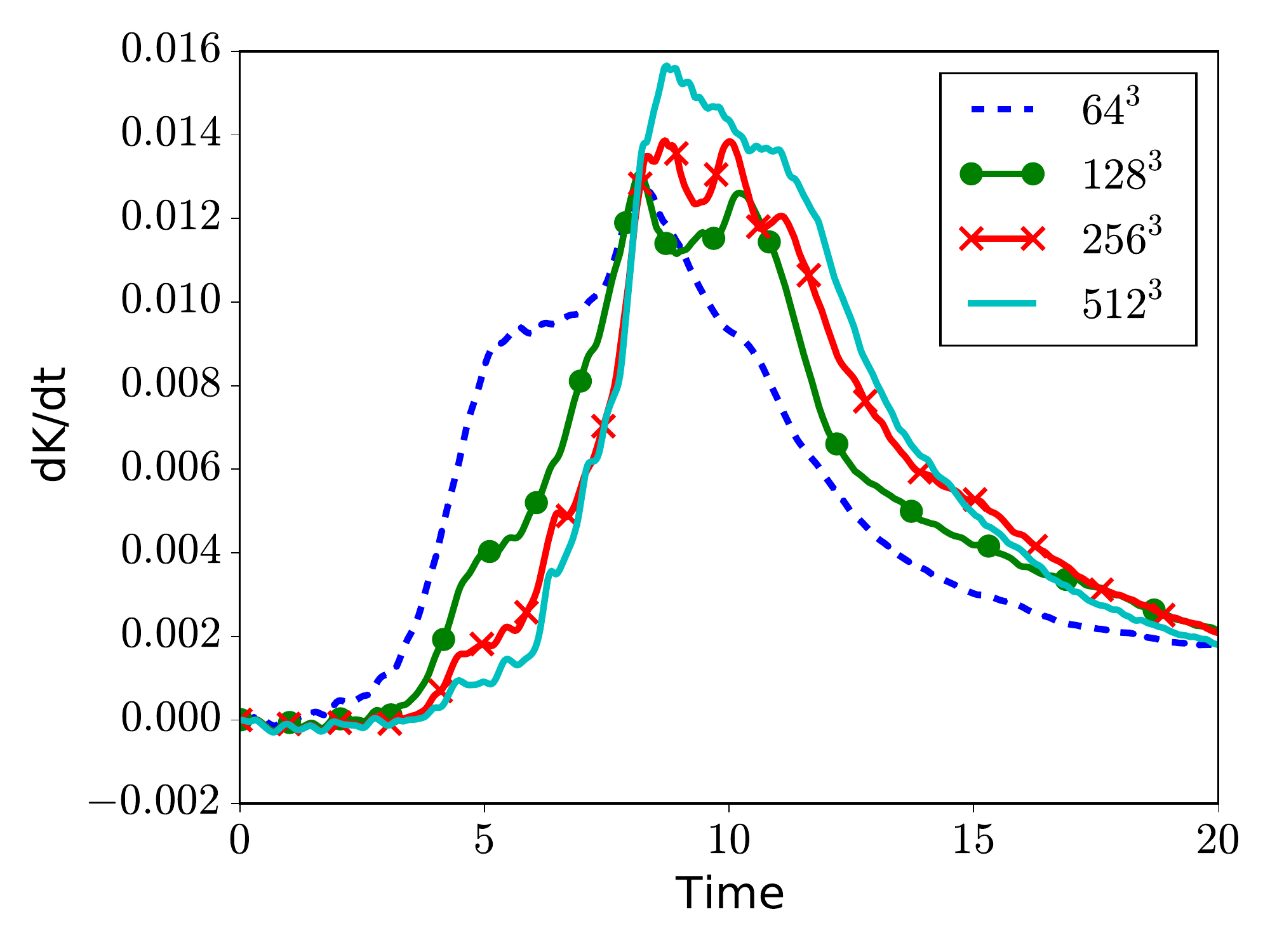}
  \caption{Decay rate of the Taylor-Green vortex, for different grid sizes.}
  \label{fig:TGV_RES}
\end{figure}

We now investigate the evolution of the kinetic energy in detail. We carry out a series of calculations at grid sizes of $64^3$, $128^3$, $256^3$ and $512^3$, until $t=20$. We carry out these calculations with $\mathrm{CFL}_{\mathrm{hydro}} \leq 10$. This choice of time-step restriction is chosen so that the kinetic energy is converged with respect to the time-step, and results in a shorter run-time than the other choices considered. We explicitly calculate the rate of change of kinetic energy density ($\mathrm{K} = \frac{1}{2}\rho\vec{v}^2$)for each grid size at each time-step.
\begin{table}
\caption{Power law decay constants fitted to the observed kinetic energy from $256^3$ Taylor-Green vortex calculations. Errors correspond to $\pm \sigma$.}
\label{table:tgv_table}
\centering
\begin{tabular}{ c  c  c}
\hline\hline
Run Name & Decay Constant  & Decay Constant \\
 & $\left(8.4 \leq t \leq 10.0\right)$ & $\left(t > 10.0\right)$ \\
\hline
TGV0.05 & $1.26\pm0.01$ & $1.828\pm0.002$ \\
TGV10 & $1.27\pm0.02$ & $1.865\pm0.004$ \\
TGV50 & $1.28\pm0.05$ & $1.92\pm0.01$ \\
\hline
\end{tabular}
\end{table}
We first compare the $64^3$ calculation shown in Fig.~\ref{fig:TGV_RES} with
those shown in Fig. 4 of \cite{viallet16}. \cite{viallet16} show that
\MUSIC\ is able to produce consistent results for a range of Mach numbers,
$10^{-1} \leq M_s \leq 10^{-6}$. However results presented here show
fluctuations around the profile presented in \cite{viallet16}. As these
fluctuations only manifest in \MUSIC\ simulations with $M_s > 10^{-1}$ they
are likely a result of acoustic fluctuations. Similar fluctuations are also present in Fig. 2(e) of DFGY2007. They are not present in the incompressible conventional LES calculations presented in DFGY2007.

In ILES calculations non-ideal effects should become less influential with increased grid size. Therefore as the grid size is increased in ILES calculations the solution should tend towards higher Reynolds number results from conventional DNS calculations. We initially compare the evolution of kinetic energy from \MUSIC\ with Fig. 2(a) of DFGY2007, which shows results from the DNS calculations of \cite{brachet1983small}. The peak dissipation is observed around $t=9$ for all \MUSIC\ calculations. This is also seen in all DNS calculations shown in DFGY2007, except the lowest Reynolds number, $\mathrm{Re} = 400$, which shows a broad peak, from around $t=6$ to $t=9$. Such a period of high dissipation is also observed in the lowest grid size \MUSIC\ simulation $64^3$, albeit with an additional peak at approximately $t=9$.

Two general patterns of behaviour can be observed with increasing grid size in Fig.~\ref{fig:TGV_RES}. Firstly, the initial high rate of dissipation around $t=5$ quickly reduces with increasing grid size. This is observed both in the DNS calculations of \cite{brachet1983small}, as well as in the MILES calculations shown in figure 2(e) of DFGY2007. Additionally, the maximum dissipation observed at $t=9$ increases with increasing grid size. A similar pattern is seen with increasing Reynolds number for DNS calculations. The peak value of dissipation in the $512^3$ \MUSIC\ calculations appears comparable to that observed in DNS calculations with Reynolds numbers of $3000$ and $5000$.

Finally we note the double peak feature in the dissipation rate, seen for both $128^3$ and $256^3$ grid sizes in \MUSIC\ calculations. Such a double peak is also apparent in the $128^3$ MILES calculations of Fig. 3 in DFGY2007 (calculated using the TURMOIL3D code, \cite{youngs1991three}), but not other MILES calculations. DFGY2007 suggest that such a double peak feature could be produced by more dispersive numerical schemes. However, DFGY2007 do not show whether this peak is present in other calculations using TURMOIL3D, so further comparison is not possible.

We have demonstrated not only the capability of \MUSIC\ to reproduce features of the decay of the Taylor-Green vortex seen in other ILES calculations, but also that an increase in grid size reproduces the same qualitative changes in dissipation seen in DNS calculations of increasing grid size. We stress that this work is not in itself a verification of the ILES paradigm. We do show that whilst an increase in the computational time-step does result in fluctuations of the observed kinetic energy, the range of these fluctuations is within the range observed for differing ILES schemes.

\section{Hydrostatic equilibrium under realistic stellar conditions}
\label{Sec:hse}
We perform a final test based on hydrostatic equilibrium under realistic stellar conditions. The \MUSIC\ code is primarily devoted to studying fluid processes in stellar interiors on timescales where hydrostatic equilibrium prevails.  It is thus crucial to verify the ability of the code to converge towards a state of hydrostatic equilibrium in a multi-dimensional configuration. As \MUSIC\ uses a staggered grid, a balance between the pressure gradient and the gravitational forces should be obtainable without resorting to more specialised methods, for example a well-balanced technique (e.g. \ct{kappeli2016well}) as used in codes with co-located variables.

The stellar model selected for this test is a 20 $\msol$ Main Sequence star with zero metallicity calculated with the Lyon one-dimensional (1D) stellar evolution code \citep{baraffe91, baraffe98}. The 1D model used as an initial setup for the present test is characterised by a surface luminosity $\mathrm{L} \sim 1.9\cdot10^{38}\,\mathrm{erg}\,\mathrm{s}^{-1} \left(\sim 5\cdot10^4\,\mathrm{L}_\odot\right)$, radius $\mathrm{R} \sim 1.9\,\mathrm{R}_\odot$ and effective temperature $T_\mathrm{eff} \sim 6.2\cdot10^4\,\mathrm{K}$. It is in thermal equilibrium, meaning that the nuclear energy production in the central regions counterbalances the energy loss at the surface.
We chose this model because of its simple interior structure, with a convective core and a radiative envelope. Due to the absence of metals in the envelope this model exhibits low radiative opacities in the outer layers. Consequently convection is not able to develop close to the stellar surface, and we are able to choose a fully radiative portion of the stellar envelope for our numerical domain.

\begin{figure}[h]
\centering
\includegraphics[width=7cm, trim = 0cm 5cm 0cm 4cm, clip=true]{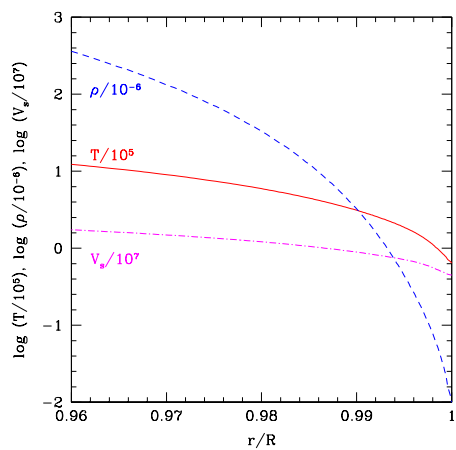}
  \caption{Radial profiles from the 1D model of the temperature (in units of $10^5\mathrm{K}$), density (in units of $10^{-6}\mathrm{g}\,\mathrm{cm}^{-3}$) and sound speed (in units of $10^7\mathrm{cm}\,\mathrm{s}^{-1}$) in the outer radiative envelope of a 20 $\msol$ star with zero metallicity. $\mathrm{R}$ is the total stellar radius.}
\label{struc}
\end{figure}

The test is performed in two-dimensional spherical geometry (with azimuthal symmetry) that considers only a small portion of the radiative envelope.
In order to obtain rapid convergence whilst using large CFL numbers, we avoid the region very close to the surface characterised by steep temperature and density gradients (see Fig. \ref{struc}).
We use a grid size of $120\times120$. The radial grid has a fixed radial spacing and is defined between $0.96\,\mathrm{R}$ and $0.99\,\mathrm{R}$. In the angular direction, the grid covers the region $50^\circ \leq \theta \leq 55^\circ$.

Periodic boundary conditions in the angular direction are used. The boundary
conditions at the radial extent of the domain are reflective for the radial
velocity component, and stress-free for the tangential component. The inner
and outer radial boundary conditions on the energy flux assume the constant
luminosity given by the 1D initial model. The inner and outer radial boundary
conditions for the density are based on the assumption of hydrostatic
equilibrium \citep[see Eq. (5) of][]{pratt16}. \cite{pratt16} tested various
boundary conditions and this set provides the best convergence toward hydrostatic equilibrium measured by the  maximum velocity magnitude obtained at the end of the simulation.

\begin{figure}[h]
\centering
\includegraphics[width=6cm, trim = 0cm 5cm 0cm 4cm,
clip=true]{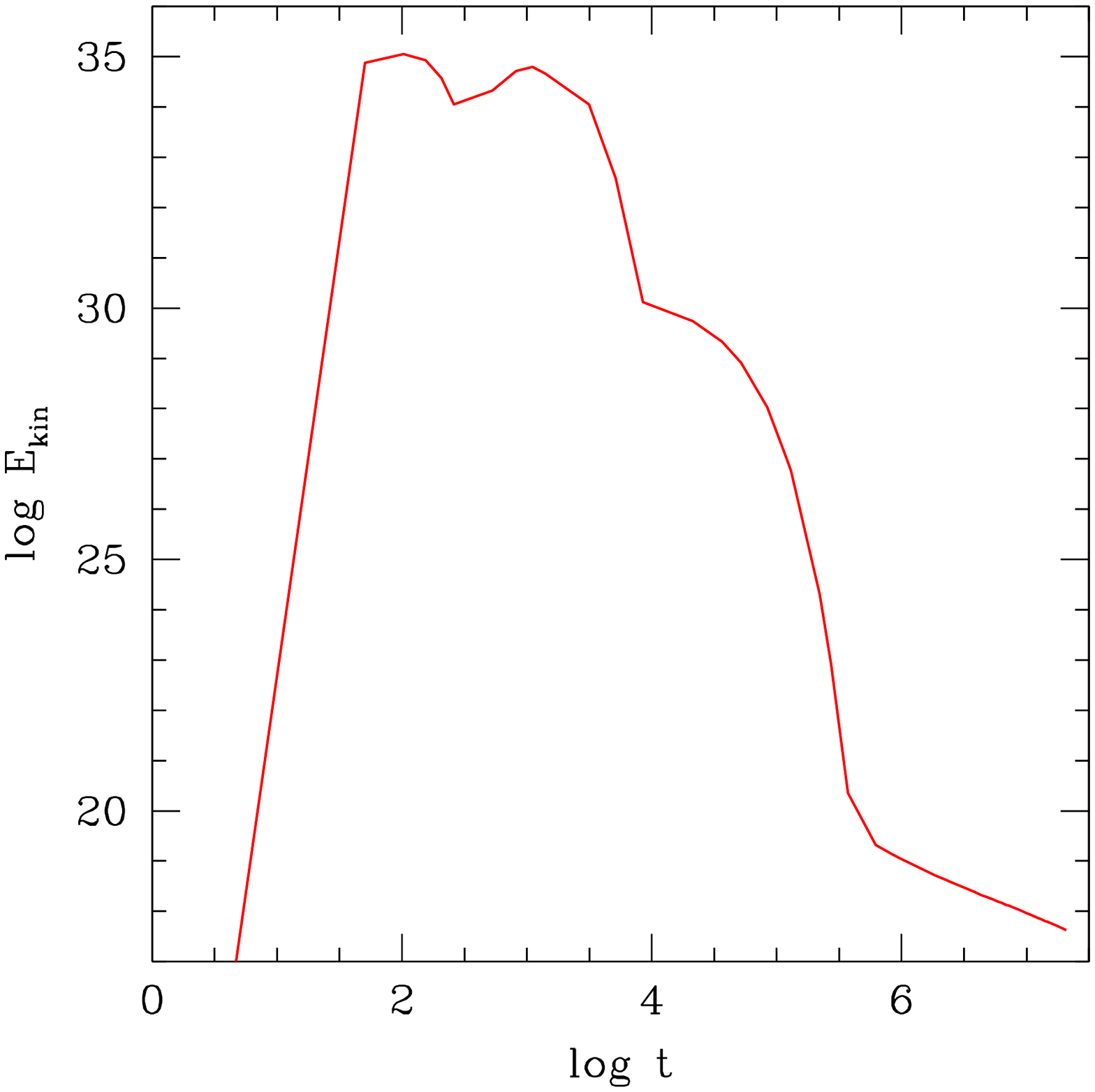}
  \caption{Evolution of the total kinetic energy $E_{\rm kin}$ (in ergs) during the relaxation process toward hydrostatic equilibrium in the stellar model. Time $t$ is in seconds.}
\label{Ekin}
\end{figure}

The model requires some time to relax toward very low velocity magnitudes that characterise the state of hydrostatic equilibrium. This is illustrated in Fig.~\ref{Ekin} by the evolution of the total kinetic energy contained in the numerical domain. After $10^6\mathrm{s}$,  the highest velocity magnitude within the domain remains around $\sim 7\cdot10^{-5}\mathrm{cm}\,\mathrm{s}^{-1}$. This low velocity corresponds to a Mach number of $\sim 10^{-11}$. The minimum value for the velocity magnitude is around $\sim 8\cdot10^{-10}\mathrm{cm}\,\mathrm{s}^{-1}$.

The most severe constraint on the timestep during the relaxation process is imposed by the radiative CFL number, defined by eq. \eqref{eq:cfl_rad}. This stems from the combination of high temperature,
low density and low opacity in the stellar model, resulting in very high radiative diffusivity $D_{\rm rad} \equiv \chi/(\rho \mathrm{c_P})  \propto T^3/(\kappa \rho^2)$, with $\mathrm{c_P}$ the specific heat at constant pressure and the other quantities defined in eq. \eqref{eq:chirad}.
We limit the radiative CFL number to 500 to reduce the number of non-linear iterations and to obtain the best performance of our solver. The preconditioner within \MUSIC\ is designed to target the physics which is restricting convergence. Due to the level of thermal diffusion in this problem it is necessary to apply the form of the physics based preconditioner which treats thermal diffusion implicitly. Without targeting the thermal diffusion with the preconditioner, the convergence of the linear system fails. The large time-step facilitated by the application of this preconditioner allows the structure to settle towards equilibrium efficiently, without the need of explicit damping. We have not tried to fine-tune the parameters of our solver \citep[see][]{viallet16} to reach lower velocities. We consider these results and the convergence toward a hydrostatic equilibrium state as satisfactory given the extremely low Mach numbers reached at the end of the relaxation process.
\section{Conclusion}
\label{Sec:conclusion}
This work builds on previous descriptions of the \MUSIC\ code by providing a
series of non-linear, multi-dimensional tests. In a model of the
Rayleigh-Taylor instability \MUSIC\ produces comparable mixing layer widths to
the well established \ATHENA\ code. The test was additionally used to assess
the new implementation of passive scalars within \MUSIC. The Kelvin-Helmholtz
test of \ct{mcnally12} provides reference solutions for peak kinetic energy,
and the mode amplitude, which are both reproducible using the
\MUSIC\ code. Furthermore the variable nature of the convergence of the
velocity field for this test problem is examined. Like many other
astrophysical codes \MUSIC\ does not include explicit viscous terms. Using the
Taylor-Green vortex the ability of \MUSIC\ to reproduce features of
established ILES codes, and conventional LES codes is shown, as well as
observations suggesting an increasing effective Reynolds number with
increasing grid size. Finally, \MUSIC\ converges towards the hydrostatic equilibrium within a radiatively dominated portion of a star, in an efficient manner through the application of a preconditioning technique adapted to such a problem.

Whilst this work aims to increase confidence in \MUSIC\ calculations, we intend it to be of general use as the basis of a code comparison test suite for hydrodynamics. Such a benchmarking exercise provides confidence and credibility to simulations. This work concludes the development of the hydrodynamical core of \MUSIC. Future work will focus on applications to stellar interiors, such as convective overshooting and shear-driven mixing.

\begin{acknowledgements}
 This project has received funding from the European Unions Seventh Framework Programme for research, technological development and demonstration under grant agreement no 320478. The calculations for this paper were performed on the DiRAC Complexity machine, jointly funded by STFC and the Large Facilities Capital Fund of BIS, and the University of Exeter Super- computer, a DiRAC Facility jointly funded by STFC, the Large Facilities Capital Fund of BIS and the University of Exeter. We are very thankful to Colin McNally for providing his results for the Kelvin-Helmholtz test.
\end{acknowledgements}

\bibliographystyle{aa.bst}
\bibliography{references}

\end{document}